\documentclass{sig-alternate}

\numberofauthors{1}

\usepackage{xcolor}
\usepackage{url}
\usepackage{hyperref}
\usepackage{paralist}
\usepackage{xspace}

\usepackage{ifthen}
\newboolean{anonymous}
\setboolean{anonymous}{false}

\ifthenelse{\boolean{anonymous}}{
  \author{author blanked for blind review}
}{
\author{
\alignauthor
Matthieu Moy\\
\affaddr{Grenoble-INP (Ensimag), Verimag UMR 5104}\\
\affaddr{Grenoble, F-38041, France}\\
\email{Matthieu.Moy@grenoble-inp.fr}
}
}

\toappear{To appear in ITiCSE 2011, the 16th Annual Conference on
  Innovation and Technology in Computer Science Education}

\ifthenelse{\boolean{anonymous}}{
  \newcommand{\ensimag}[0]{\textit{School name blanked for blind review}}
  \newcommand{\citeensiwiki}[0]{\cite{ensiwiki-blind}}
  \newcommand{\citetextbook}[0]{\cite{poly-intro-unix-blind}}
  \newcommand{\demosurl}[0]{\textit{(url blanked for blind review)}}
  \newcommand{\feedbackurl}[0]{\textit{(url blanked for blind review)}}
  \newcommand{\gitoriousproject}[0]{\textit{(url blanked for blind review)}}
}{
  \newcommand{\ensimag}[0]{Ensimag}
  \newcommand{\demosurl}[0]{\url{http://www-verimag.imag.fr/~moy/demos-unix-training/}}
  \newcommand{\feedbackurl}[0]{{\scriptsize\url{http://ensiwiki.ensimag.fr/index.php/Discussion:TP_Unix_-_Jeu_de_piste}}}
  \newcommand{\gitoriousproject}[0]{\url{http://gitorious.org/unix-training}}
  \newcommand{\citeensiwiki}[0]{\cite{ensiwiki}}
  \newcommand{\citetextbook}[0]{\cite{poly-intro-unix}}
}

\title{Efficient and Playful Tools to Teach Unix to New Students}

\begin{document}
\maketitle

\begin{abstract}
  Teaching Unix to new students is a common tasks in many higher
  schools. This paper presents an approach to such course where the
  students progress autonomously with the help of the teacher. The
  traditional textbook is
  complemented with a wiki, and the main thread of the
  course is a game, in the form of a \emph{treasure hunt}. The course
  finishes with a \emph{lab exam}, where students have to perform
  practical manipulations similar to the ones performed during the
  treasure hunt. The exam is graded fully automatically.

  This paper discusses the motivations and advantages of the approach,
  and gives an overall view of the tools we developed. The tools are
  available from the web, and open-source, hence re-usable outside the
  \ensimag{}.
\end{abstract}




\section{Introduction}

\ensimag{} is a french engineering school of computer science and applied
mathematics. The computing environment is essentially based on Unix
(servers and workstations), which the students have to learn when they
enter the school. The students therefore follow a quick unix-learning
course at the beginning of the first year.

While this introduction to Unix has sometimes been considered as an
unimportant course, we believe it is a fundamental mistake to
underestimate its importance: the
few hours taken at the beginning of the year to learn and train with
the basics influence the students' productivity for the next 3 years,
and even if learning Unix is not a goal in itself, it conditions the
success of further courses. This paper discusses the challenges and
solutions set up in the \ensimag{} the last few years to reconsider the
introduction to Unix as an important course, to motivate the students
and teach them as efficiently as possible.

After detailing the context and our motivations in
section~\ref{sec=motivations}, we give a quick overview of the course
and training material in section~\ref{sec=material}. The main
contributions of this paper are two tools used in the course:

\begin{compactitem}
\item A set of exercises in the form of a \emph{treasure hunt}, used
  by the students to train autonomously during the course (presented
  is section~\ref{sec=treasure-hunt}).
\item A \emph{lab exam} that allows grading the students at the
  end of the course, with practical manipulations (presented in
  section~\ref{sec=lab-work-exam}).
\end{compactitem}

Both tools are published as open-source software, and could be re-used
and adapted by other teachers/schools.

\section{Particularities of the Unix Introduction and Motivations}
\label{sec=motivations}

One challenging aspect of teaching Unix to beginners is the
heterogeneity of students. All of them have used a computer prior to
entering the school,
but around half never used Unix before. On the other hand, a number of
them had some exposure to user-friendly Linux distributions, and a
small number are already command-line gurus. The difficulty is to let
the course be effective to total beginners, while remaining interesting
to the other students.

Our way to tackle heterogeneity is twofold. First, we designed the
course to let the students learn at their own pace, with a maximum
degree of autonomy. This is not hard since we teach them practical
aspects first: all the classes are done in the computer rooms, one
student per machine. The students essentially follow a textbook plus
some on-line exercises, and teachers provide advices and answers to
questions.
Since the goal of the introduction is really to teach \emph{practice},
we do not enforce team work like in
e.g.~\cite{DBLP:conf/iticse/HurtadoV05}, but ask each student to work
individually. Mutual help is appreciated and encouraged, but we do not
want a situation where one student holds the keyboard, and another
watches without practicing.

We provide the students teaching material containing all the basics,
but extensively use remarks targeted to more advance users. For the most advanced users,
we provide additional documentations on various topics and pointers to
external documentations, so that they can start learning concepts that
they would otherwise miss, or learn a few months later.

A common example of advanced topic is revision
control-system.
Most users won't
understand the need for one immediately, and we wait for some time
before imposing them one. Still, encouraging a handful of students to
start using it spreads the knowledge with a network effect: these
students will encourage (or force!) their co-workers to use it.
In short: teaching useful advanced concepts to advanced users also
help beginners in the long run. Another subtle advantage of giving
material to advanced users is that it helps keeping them in the
machine rooms (as opposed to missing classes they don't
need), indirectly promoting mutual help.

The other challenging aspect is the students motivation. Most students
come from so-called ``classes preparatoires'' in the french system,
with a huge exam pressure, and many of them expect the engineering
school to be easy enough to pass exams without working.

Another issue with students' motivation is the desire to learn
Unix and the command-line, as opposed to another operating system
(like Windows or the graphical part of Mac OS).
We do believe in the pedagogical qualities of Unix and the
command-line, since they somehow \emph{force} the user to understand
what s/he's doing, but the students often don't understand the need to
re-learn the computer basics (file manipulation, launching
applications, ...), since they already know it with other paradigms on
other platforms. The vast majority of students have a personal laptop
with either Windows or Mac OS installed, and some of them wouldn't use
Unix at all unless we find the right arguments.

In addition, the organization of the school does not leave room for a
lot of teaching hours for this course. We have just 10 hours to teach
them all this. For the slowest students, this is far below what would
be needed, so we need to convince them to complement the classes with
their personal work.

We therefore needed to work a lot on students motivation. The first
thing we did was to start with very simple manipulation in a graphical
environment, to avoid scaring new students at the first contact. The
message perceive by students should be more like ``look, you can still
do the kind of things you're used to, \emph{but you can also do many
  others}'' than ``forget all you know, and then learn''. Unix and the
command-line should not be the new things to be \emph{scared} of, but
the friendly companion which will help them to learn new things. For
example, we show them how to surf the web, read their mail, and use a
word processor before diving into the command-line. Also, changing the
recommended text editor from \texttt{vi} to \texttt{emacs}, we noticed
that students started to actually use the editor we recommended them
(at the time when \texttt{vi} was recommended, the majority of
students was using \texttt{nedit}!).

After avoiding to scare students during the first contact, a lot is
still to be done to maintain their motivation during the course. One
tool for this is the ``treasure hunt'' game described in
section~\ref{sec=treasure-hunt}, that the students follow all along
the course. The basic idea is a sequence of levels, designed such that
accessing level $n+1$ requires performing a manipulation described at
level $n$. This creates a little (sane) competition between students,
and many of them really have fun while learning.

\label{motivation-exam}
After setting up this treasure hunt and rewriting the textbook, we had
very positive feedback from students and all of those who completed the
game were even thankful for the fun they had. However, less than half
of the students did complete the game. Based on this observation, it
was clear that we had to complement the course with the other side of
students motivation: exams and grading. We therefore designed an exam,
essentially based on the same manipulations as the treasure hunt, but
for which the questions are independent. We will describe it in
details in section~\ref{sec=lab-work-exam}.

\section{Basic Teaching Material}
\label{sec=material}

Before detailing the main contributions of the paper, we give an
overview of the course, and the teaching material.
The teachers do
not provide any lecture and only occasionally talks to the whole
students group: to give the instructions, and to show a few demos for
technical aspects that are better explained live than on a textbook or
webpage.

\subsection{Textbook}

The course starts with the distribution of a textbook
(\citetextbook, french only). The textbook is
made to be read linearly. It was designed to allow multiple levels of
reading: remarks for advanced users are identified visually
(technically, using \LaTeX{} macros). Each notion taught by the
textbook is illustrated immediately with a small exercise (also marked
visually to draw the attention of students).

While previous versions of the textbook used to try being independent
from the school, we decided to adapt it deliberately to the school and
the course, making frequent references to the school's intranet, the
particular configurations of the machines they are using\dots{} While
this requires more effort from the teachers to maintain the book up to
date, we believe this gives a real added value over a random Unix
introduction found on the Internet.

Having the textbook in
paper form makes reading long text more comfortable than on-screen
reading, and the linear structure ensures everybody goes through all
the important points.

\subsection{EnsiWiki: Students' and Teachers' Wiki}

In complement to the textbook, a wiki called EnsiWiki~\citeensiwiki
is provided to the students. Historically, it is a merge of a
wiki launched by the school and an independent initiative carried out by
students. We try to maintain the equilibrium between teachers'
and students' contributions (both having full write
access).

As opposed to the textbook, the wiki is not meant to be linear. It
doesn't have a beginning and an end, but is basically a set of pages
with hyperlinks (plus a classification using the category system of
MediaWiki).
%
Students read pages that are of interest
to them in the order they wish, and of course, add and improve pages
as they wish: it's a wiki! Unlike the textbook, the wiki is not just a
starting point, but will really accompany the students throughout
their studies. It is public, and indexed
by web search engines, so searching for information is usually
relatively easy.

The duality between the textbook and the wiki can be summarized as
follows: as a beginner, the textbook tells the students what they have
to learn, but when the students know what they are looking for, the
wiki should be able to provide them the information needed.

A positive side-effect of the wiki is that it increases the visibility
of the school on the web. Some of the articles are of great quality
(including many articles written by students!), and are very well
ranked on popular search engines.

\section{Treasure Hunt}
\label{sec=treasure-hunt}

\subsection{Principle}

As discussed above, the real challenge in this course is not to
provide content to students, but to motivate them, and to make sure
they work autonomously but efficiently. One tool we developed to
accomplish this goal is the \emph{treasure hunt} (called ``jeu de
piste'' in its original version, since the course is in french).

The principle is simple, basically an electronic version of the
children's game: the first level contains instructions to reach the
second, which itself contains instructions to get to the third, and so
on. It contains 28 levels (plus 7 bonus levels to make sure the
geekest students---and teachers---to have fun too!).
It can be seen as a Unix-ish, pedagogical version of
ouverture-facile~\cite{ouverture-facile}.

In theory, this is similar to a set of unrelated exercises, but in
practice, this makes a real difference, with at least the following
advantages:

\textit{The students cannot mistakenly think they completed the
  exercise}. Either they solved the level, and know it, or they didn't,
  without half-measure. This is a key point to allow autonomous work.
\textit{The students cannot skip an easy exercise}. When practical
  manipulations are proposed in the textbook, they are easily
  overlooked as too easy, and skipped. This can result
  is a false impression of having completed the work with the
  reasoning ``I went through the textbook, that's all too easy for me,
  I didn't need to do the exercises''.
\textit{The students cannot skip a hard exercise}. Some levels are
  purposely hard, and almost unfeasible by
  beginners without help. The rational is twofold: first, this
  encourages the students to help each other (the game itself is not
  graded, we ask the students not to give answers directly, but
  cooperation is welcome), and second, it force students to ask for
  help to the teacher. Autonomous work doesn't mean teacherless work:
  students go through the game at different speeds, but the teacher is
  indeed very active to answer questions.
\textit{Off course, this makes the sequence of exercises funnier than
  traditional ones}. Students are usually looking forward to reaching
  the last stage, and reading the textbook is a mandatory step to reach
  this. Not all students enjoy the fun of the game (at least, not all
  of them have as much fun playing the game than I had \emph{creating}
  it!), but on average, the effect on motivation was very positive.

Note that these advantages come with a risk: students blocked at one
level would miss the end of the game. It is the teacher's role to make
sure this does not happen, by advising students, and sometimes by
making surveys (who started the treasure hunt? Who went past level
$X$?\dots).

\subsection{Contents of Levels}

The nature of the game requires the instructions for each levels to be
hidden, and only discoverable by following instructions. We use
essentially two kinds of tricks to achieve this:

\begin{compactitem}
\item Instructions obfuscated with simple encryption schemes,
  typically variants of rot13. The text is easily
  available, but can be deciphered only with the
  instructions.
\item Instructions in a file, in a non-listable directory. Files are
  either in the filesystem of a machine the students have access to
  (the Unix permission \verb|--x| on directories allows giving access
  to files when user know their names), or on a website with directory
  listing disabled (so, students can easily access a level when they
  know its URL only).
\end{compactitem}

The game follows the chapters of the textbook. Following are some
examples of levels:

\renewcommand{\subsubsection}[1]{\textbf{#1:}}
\subsubsection{Internet}
The game starts with a rot13-encoded piece of text. The player is told
that rot13 is used, without being told it is. The expected solution
is basically to search the web, and find, e.g. \url{http://rot13.com},
which allows online decyphering very quickly. As with many other
levels, the solution of the level gives a few comments on the solution
and the way to find it. In this case, the text insists on the need for
students and future engineers to be able to quickly find the
information.

Next steps include some navigation in the wiki, a script sending an
email to the students, so that they are forced to read their email.

\subsubsection{Basics}
In this section, the students must copy a file from another user's
directory. The file is an obfuscated Ada (the language taught in
first year in the \ensimag{}) source-code that must be
compiled and executed to provide instructions.

\subsubsection{Useful Applications}
This section consists in opening files made for various applications.
Students have to compile a \LaTeX{} file, open an OpenOffice.org file
and a PNG image.

\subsubsection{Text Editor}
Again, students are provided Ada source code. This time, the file is
very long, and contains a few syntax errors. Being familiar with a
text editor (typically, being able to jump to a given line number) is
almost mandatory. Then, another piece of code is given to the
students, but split into 3 pieces, in a text file, within the
instructions, and in an OpenOffice.org file, to force students to do
inter-applications cut-and-paste.

\subsubsection{Commands and Tools, and Bash}
These two chapters are key ones, where students learn the essentials
of the command-line. Players have to use a few commands like
\texttt{file}, \texttt{grep}, \texttt{find}, \texttt{sort},
\texttt{diff}, \texttt{tar}, \texttt{chmod}, find hidden files, play
with input/output redirects (\verb+|+, \verb+<+ and \verb+>+) and
wildcards.

The hardest level consists in finding the biggest file within a
directory (and its sub-directories). Students usually need the help of
their teacher, which gives a good opportunity to explain or re-explain
the concepts of pipe and the \texttt{xargs} command, with a solution
along the lines of ``\texttt{find . -type f | xargs wc -c | sort -n |
  tail -n 2}''. Not all students really understand the complete
command-line, but exposing them once to a complex command gives them a
hint on what it's possible to do with Unix once they master it, and
therefore what they would lose by not learning it.

\subsubsection{Remote access}
This chapter provide a few ways for the students to access a machine
remotely, trying to answer the common question ``how can I work with
the \ensimag{}'s machines and my personal laptop'' with tools like SSH.
Students have to fetch a file from a remote machine with
\texttt{sftp}, and to execute remote commands with \texttt{ssh -X}.

\subsubsection{Bonus Levels}
This section is presented to students as non-mandatory. Beginners are
not supposed to be able to solve all levels when they enter the
school, but should be able to do so after a few months. Levels include
finding information in HTTP headers of a webpage, basic
shell-scripting, navigating in the history of a directory managed by
the Git revision-control system,
using \texttt{strace} or
navigating in the \texttt{/proc/} virtual filesystem, and using SSH
private/public keys. The last level gives a pointer to the source code
of the generation scripts. Students are encouraged to contribute new
levels (but none actually did up to now).

\subsection{Generation Library}

The complete set of scripts used to generate the treasure hunt is
available on the web, and is open-source. The instructions given to
students are in french, but the source code is written and commented
in english, with a relatively clean separation between library code
and the actual code for each level, so it can easily be adapted to
other schools, in other languages.

Technically, the library provides a few source code obfuscation
functions (to generate unreadable \LaTeX{}, Ada or C code), and plain
text encoding/decoding. For example, script generating the level about
input/output redirects takes the instructions for the next level,
encodes it, and provides the students a decoder that will read the
encoded instructions on its standard input.
It can be found online at the following URL:
\gitoriousproject

\subsection{Students Feedback}

The feedback from students completing the treasure hunt can be found
on
EnsiWiki\footnote{\feedbackurl}.
Only 101 students out of 210 provided feedback. We have unfortunately
no way to distinguish students who did not provide feedback because
they didn't take the time to do it, and ones who didn't because they
did not complete the game (the next version of the game will detect
automatically when students reach some levels). The result is probably
biased towards positive feedback.

Still, the majority of comments are \emph{very} positive. For example,
we can count 6 occurences of ``thank you'', which is in our
experience seldom used in students feedback. The most frequently used
words include ``good'', ``friendly'', ``playfull'', ``instructive'', ...
Many students confirm that they did manage to complete the game
without having prior knowledge about Unix.

Since the starting point of the game is public, 2nd year and 3rd year
students have access to it too. We had several requests to install the
necessary files on the servers they use, because they wanted to play,
too! Some students even offered to host the game on their club's
server.

\section{Lab Exam}
\label{sec=lab-work-exam}

Despite the very positive feedback we got from students actually
completing the textbook and treasure hunt, this turn\-ed out to be
insufficient to motivate \emph{all} the students to actually complete the
work. We have therefore set up an exam, designed to be very easy
for anyone having done the work seriously, and very hard
otherwise. Since the whole course is done in computer-rooms, it would
make no sense to have a theoretical exam, so the exam is also done on
computers, and consists of a set of technical manipulations.

\subsection{Principle}

The exam is made of a set of questions, highly inspired from the
treasure hunt. The questions are independent: instead of giving access
to the next level, the manipulation asked provide a key, that is used
to answer questions on a web interface.

\label{sample-questions}
To fix the ideas, the first (simple) question is ``The answer for this
question is in the file c73df134.txt in your working directory (it is
a text file).''. The
student's account contain a file named \texttt{c73df134.txt}, whose
content is ``The answer is 3d61f5e5'', and the students must copy the
string \texttt{3d61f5e5} in a web-interface to validate the answers.

The design was done with the following ideas in mind:

\textbf{Automatic grading:} the exam was set up to force the students'
  work, but should not overload the teachers. Setting up the exam was
  a rather large one-time effort, but grading should be as simple as
  executing a script to collect the answers.

\textbf{Immediate feedback:} the obvious problem with automatic grading
  is that automatic tools do not distinguish ``almost correct
  answers'' and ``actually correct'' ones. To solve this issue, the
  student get an immediate feedback: when giving a correct answer, the
  question is validated as correct, and otherwise, the students get an
  unlimited number of retries. This implies that the answers of the
  questions have to be impossible to find by trial-and-error.

\textbf{Hard cheating:} students are close to each others in the
  computer rooms. To avoid easy cheating, the exam is generated on a
  per-students basis. For almost all questions, the answer is
  different from a student to the next, even though the technical
  manipulation required to obtain it is the same. Questions are sorted
  in a pseudo-random order (which is possible since questions are
  independent). Also, during the exam, the machine's network is
  restricted with a firewall.

\textbf{Simple technologies:} since it is used for grading, the
  robustness of the exam infrastructure is critical. Also, we wanted
  the infrastructure to be re-usable outside the school. We have
  therefore chosen the simplest technologies to accomplish this, with
  a rather unix-ish design: shell-scripts generate the exam, and the
  web-interface used during the exam is simple PHP+SQL scripts (tested
  with both MySQL and PostgreSQL). No JavaScript, no external
  dependencies.

At the beginning of the exam session, the machines are initialized
with an account containing only the files needed for the exam. During
the exam, the students will have to perform manipulations on this set
of files, and will validate the answers through a web-interface (a
single web-page showing a text-box and a submit button for each
question). The answers are stored in a database, and the grades are
extracted from this database at the end of the exam.

\label{sha1sum}
To differentiate the answers for each students, the answers are
pseudo-random (typically looking like \texttt{3d61f5e5}). Actual
random would be possible, but with the great drawback of being
non-reproducible. Cryptographic hash functions~\cite{hash-functions}
provide an
elegant solution to this: we compute the answer to each question as
the sha1 sum of the student's login concatenated with the name of the
question, and a secret key for each exam. This way, regenerating the
exam several times yields the same answers each time (which can be
crucial if something goes wrong and the exam has to be regenerated at
the last minute...).

\subsection{Content of the Exam}

The exam contains 28 questions where students are asked to compile Ada
and \LaTeX{} code, to find files in a directory containing hundreds of
subdirectories and files, to extract zip, tar and gz compressed
files, to play with input/output redirects, find the size of a file,
the destination of a symbolic link, to use \texttt{sort},
\texttt{grep}, \texttt{diff}, \texttt{kill}, to use Control-z to
suspend a running executable, to download a file with \texttt{sftp},
to connect to an account with \texttt{ssh}, ... The exam is feasible
in 30 minutes by an expert user, and we let 1 hour to the students, so
that most of them do not have time to reach the end. This way, we test
students on their speed as well as their skills.

\subsection{Demo Mode}

During the exam, users are identified with the IP address of their
machine, and answers are recorded in a database. We made a variant of
the generation scripts that do not use any authentication, and stores
answers in PHP session variables (i.e. simple storage based on browser
cookies). In this mode, the students can practice with a few
questions, getting the same interface as the real one, but their
answers are not transmitted to teachers. Such a demo was put online
during the course, with trivial questions to get used to the web
interface, and a few questions extracted from the actual exam. Some
examples are available online\footnote{\demosurl}.
Next year, this demo will be integrated as one level of the treasure
hunt.

\subsection{Generation Library}

As for the treasure hunt, the technical infrastructure behind the
lab exam is published as Open Source. We did not publish the full
set of questions, to avoid students finding it and publishing
ready-made solutions, but this can be distributed in private upon
request.

The generation library consists in a set of shell-scripts functions.
The user defines two shell functions per questions. These functions
will be called once per student. One gives the question, as will be
displayed to the student (possibly depending on the student's login),
and the other sets up the files as will they will be stored on the
student's account. The first argument to this function is the expected
answer. For example, the trivial question mentioned
in~\ref{sample-questions} is implemented as:

{\small
\begin{verbatim}
desc_question_text () {
    echo "The answer for this question is in the file
<tt>$(hash textfile).txt</tt> in your working directory
(it is a text file)."
}

gen_question_text () {
    echo "The answer is $1" > $(hash textfile).txt
}
\end{verbatim}
}
Notice the use of \texttt{hash} to compute the file name. It is
provided in the generation library, and implements the pseudo-random
based on \texttt{sha1sum} described in section~\ref{sha1sum}.
The same code-obfuscation library as the treasure-hunt is used.

The execution of this script will provide a directory containing one
subdirectory per student, with the content of their account (then,
other mechanisms have to be used to deploy it on students machine),
and an SQL file to initialize the database with questions, and
expected answers. We also provide the PHP files needed for the web
interface during the exam.

The generation library, and a heavily commented example of exam that
serves as documentation, can be found at the following URL:
\gitoriousproject

\section{Conclusion}
\label{sec=conclusion}

We presented the ``introduction to Unix'' class of the \ensimag{}.
Starting from a relatively standard content based on a textbook, we
introduced a wiki, and then two novel tools: the treasure hunt allows
learning autonomously in a playful way, and the lab exam
ensures the most recalcitrant find a motivation to complete the
work.

In the past, we have already set up several lab exams,
essentially in programming where the delivery is a program. This one
differs from the others in that the skills tested are purely
practical, and indeed relatively basic. Hence, we ask various small,
independent manipulations. Previous exams have
been very successful at motivating/forcing the students work.
Prior to this, the practical skills were tested on team work, and many
students were relying on their teammates. We believe to have made it
much harder for them to fall through the cracks, and generally feel
that students are becoming more comfortable with our computing
environment.

We do not have students feedback other than their grades as far as the
exam is concerned, but the treasure hunt received a very warm feedback.
We could already feel the effect of the exam on the presence of
students during the classes: the last two ones are non-mandatory, and
only 10 to 15 students attended them last year, compared to about 100
(i.e. half) this year! The grades for the exam were surprisingly good.
18\% of students got all answers correct, and the average grade was
15.2/20.

The technical infrastructure we developed can be compared to
Linuxgym~\cite{DBLP:conf/iticse/Solomon07}, which is based on a
complete, modified, Linux server on which the students log in to get
problems to solve. The focus of Linuxgym is scripting, while our goal
is an introduction to day-to-day use of Unix. We believe the playful
aspect of the treasure hunt, and the fact that the hunt is done
directly on the student's machines makes it more motivating and more
concrete for a first contact, but we will evaluate Linuxgym for the
more advanced Unix courses in the school.

The content of the course was tailored for an introduction to Unix.
The principle clearly does not apply to theoretical courses, and is
probably not applicable as-is in programming courses: both the
treasure hunt and the exam take advantage of the fact that each
manipulation can be solved in a few minutes, while most interesting
programming problems would take hour(s). Still, the concept can
probably be adapted to other classes involving practical aspects
(network courses would be nice candidates).

\bibliographystyle{abbrv}
\bibliography{unix-course}
\end{document}